\documentstyle[aps,prl,multicol,amssymb,amsfonts]{revtex}

\title{\rightline{ \small DAMTP-1999-112 \normalsize} \rightline{
\small quant-ph/9909016 \normalsize}
\rightline{} \centerline{Non-local Correlations are Generic in
Infinite-Dimensional Bipartite Systems} }

\newcommand{\abs}[1]{|#1|}
\newcommand{\norm}[1]{\| #1\|}
\newcommand{\trnorm}[1]{\| #1 \| _{T}}
\newcommand{\alg}[1]{\mbox{$\mathfrak{#1}$}}
\newcommand{\trace}[1]{\mbox{$\mathrm{Tr}$}(#1)}
\newcommand{\hil}[1]{\mbox{$\mathcal{#1}$}}
\newcommand{\dimn}{\mbox{$\mathrm{dim}$}}
\newcommand{\ket}[1]{| #1 \rangle}

\begin{document}

\author{Rob Clifton}
\address{Departments of Philosophy and History and Philosophy of
Science, \\ 10th floor, Cathedral of
Learning, University of
Pittsburgh, \\ Pittsburgh, PA\ 15260, U.S.A. \\ email: rclifton+@pitt.edu}
\author{Hans Halvorson}
\address{ Department of Mathematics, 301 Thackeray Hall \\
\vskip 5pt
and \\
\vskip 5pt
Department of Philosophy, 1001 Cathedral of Learning \\University of
Pittsburgh, Pittsburgh, PA\ 15260, U.S.A. \\ email: hphst1+@pitt.edu }
\author{Adrian Kent}
\address{
Department of Applied Mathematics and Theoretical Physics, University of
Cambridge,\\
Silver Street, Cambridge CB3 9EW, U.K. \\ email: a.p.a.kent@damtp.cam.ac.uk }

\date{September 3, 1999}
\maketitle

\begin{abstract}  It was recently shown that the nonseparable density
operators on the Hilbert space $\hil{H}_{1}\otimes \hil{H}_{2}$ are
trace norm dense if
  either factor space has infinite dimension.  We show here that non-local
  states---i.e., states whose correlations cannot be reproduced by any
  local hidden variable model---are also dense.  Our constructions
  distinguish between the cases $\dimn \hil{H}_{1} = \dimn \hil{H}_{2}
  = \infty$, where we show that states violating the CHSH inequality
  are dense, and $\dimn \hil{H}_{1} < \dimn \hil{H}_{2} = \infty$,
  where we identify open neighborhoods of nonseparable
  states that do not violate
  the CHSH inequality but show that states with a subtler form of
  non-locality (often called `hidden' non-locality) remain dense.
\end{abstract} \draft

\begin{multicols}{2}
\section{Introduction}

The observables of a bipartite quantum system are represented by the
set of all self-adjoint operators on the tensor product of two Hilbert
spaces $\hil{H}_{1}\otimes \hil{H}_{2}$, whose dimensions we shall
denote by $d_{1}$ and $d_{2}$, taking $d_1
\leq d_2$ without loss of generality.  It is well-known that when $d_{1}
\geq 2$ the states of
the system can be nonseparable, and it is this possibility that much
of the new technology associated with quantum information and
computation theory relies upon.  Prompted by concerns about whether
the very noisy mixed states exploited by certain models of NMR quantum
computing are truly nonseparable \cite{NMR1,NMR2}, detailed
investigations have shown that, whenever $d_{2}<\infty$, there is
always an open neighborhood of separable states surrounding
the maximally mixed state $(d_{1}d_{2})^{-1}I\otimes I$ \cite{horo,caves,zguy}.

Complementing these results, two of us~\cite{clif} have recently shown
that if $d_{2}=\infty$, the set of nonseparable states is dense, and,
therefore, there can be no open neighborhood of separable states in
that case.  It was then conjectured~\cite{clif} that the same density
result ought to hold for states which violate some Bell inequality, at
least in the case $d_1 = d_{2}=\infty$.  This does not follow
immediately from the main theorem in \cite{clif}, since the
nonseparability of a mixed state (in contrast to the pure case
\cite{gisin,rohr}) is not known to imply that it
violates a Bell inequality or that its correlations cannot be
reproduced by a local hidden variables model.  No counterexample is
known either; however, Werner \cite{wern} has shown that a local
hidden variables model can reproduce the correlations of a
nonseparable mixed state for single \emph{projective} measurements on each
component system.

We show here that the conjecture made in \cite{clif} is true.
More precisely, we show that a bipartite system
possesses a dense set of states violating the CHSH inequality for
projective measurements if and only if $d_{1}=d_{2}=\infty$,
and that the system possesses a dense set of states with
non-local correlations if $d_{1}<d_{2}=\infty$.
In the second case, we demonstrate that the states have non-local
correlations for sequences of projective measurements: we do
not exclude the possibility that they also violate a `higher order' Bell
inequality \cite{pit,peres,gisin1} involving more than two measurement
choices for each
component system,
nor do we exclude violations which involve positive operator valued
measurements.  Our results also yield an elementary proof of the main
result of~\cite{clif}.

\section{Preliminaries}
We first establish some basic facts about nonseparability and
non-locality necessary for the sequel.

Let $\alg{B}(\hil{H}_{1}\otimes \hil{H}_{2})$ denote the set of all
(bounded) operators on $\hil{H}_{1}\otimes \hil{H}_{2}$, and let
$\alg{T}\equiv \alg{T}(\hil{H}_{1}\otimes \hil{H}_{2})$ be the subset
of (positive, trace-$1$) density operators.  Throughout, we shall
consider $\alg{T}$ as endowed with the metric (and corresponding
topology) induced by the trace norm, $\trnorm{A}\equiv
\trace{(A^{*}A)^{1/2}}$.  We reserve the notation `$\norm{A}$' for the
standard operator norm.  An operator $A$ is called a
\emph{contraction} if $\norm{A}\leq 1$.  We denote the self-adjoint
contractions acting on a Hilbert space $\hil{H}$ by
$\alg{B}(\hil{H})_{s}$.  The metric induced by the trace norm is
appropriate physically for measuring the distance between quantum
states, because \cite[p. 46ff]{schat}
\begin{eqnarray}
\trnorm{D-D'} &=&\sup \Bigl\{ \abs { \trace{DA}-\trace{D'A} }:  \nonumber \\
&& \qquad \qquad A\in \alg{B}(\hil{H}_{1}\otimes \hil{H}_{2})_{s} \Bigr\}
\label{norms} \end{eqnarray}
which implies that trace norm close states dictate close
probabilities for the outcomes of measuring any observable.

For $D\in \alg{T}$, $D$ is said to be a \emph{product state} just in
case there is a $D_{1}\in\alg{T}(\hil{H}_{1})$ and a $D_{2}\in
\alg{T}(\hil{H}_{2})$ such that $D=D_{1}\otimes D_{2}$.  The
\emph{separable} density operators are then defined to be all members
of $\alg{T}$ that may be approximated (in trace norm) by convex
combinations of product states~\cite{wern}.  In other words, the
separable density operators are those in the closed convex hull of the
set of all product states in $\alg{T}$.  By definition, then, the set
of \emph{non}separable density operators is open.

 Let $A_{1},A_{2}$ be self-adjoint contractions in
$\alg{B}(\hil{H}_{1})_{s}$, and, similarly, let
$B_{1},B_{2}\in\alg{B}(\hil{H}_{2})_{s}$.  Then the corresponding operator
\begin{equation} R\equiv\frac{1}{2} \Bigl( A_{1}\otimes
(B_{1}+B_{2})+A_{2}\otimes
  (B_{1}-B_{2}) \Bigr)  \end{equation}
is called a \emph{Bell operator} for
the system $\hil{H}_{1}\otimes \hil{H}_{2}$.  Fix a density operator
$D\in \alg{T}$.  We can then define the \emph{Bell coefficient} $\beta (D)$ of
$D$ by \begin{eqnarray}
\beta (D) &\equiv &\sup \Bigl\{ \abs{ \trace{DR} }: \nonumber \\
& & \qquad \quad R\; \mbox{is a Bell operator for} \;
\hil{H}_{1}\otimes
\hil{H}_{2} \Bigr\}  . \end{eqnarray}
Bell's theorem, as elaborated by
Clauser-Horne-Shimony-Holt\cite{belltheorem1,belltheorem2},
implies that for any state $D$ and Bell operator $R$, a local hidden variable
model of $D$'s correlations  is committed to predicting
the \emph{CHSH inequality} $\abs{\trace{DR}}\leq 1$.
On the other hand, there are
always states $D$ for which $\beta (D)>1$.  We say such states are
\emph{CHSH violating}.

Convexity arguments entail that $\beta (D)$ is in fact equivalent to
the supremum taken over all Bell operators where $A_{i},B_{i}$ are
self-adjoint \emph{unitary} operators satisfying
$A_{i}^{2}=B_{i}^{2}=I$, i.e., generalized spin
components \cite[Prop.~3.2]{summ}.  For completeness,
we set out a detailed proof of this fact in
Appendix~\ref{app}.  Unless otherwise noted, we
henceforth assume that all our Bell operators are constructed
out of self-adjoint unitaries.  Moreover, for such Bell operators we always
have \cite{land}
\begin{equation} \label{eq:landau} R^{2}=I\otimes
  I-\frac{1}{4}[A_{1},A_{2}]\otimes [B_{1},B_{2}],
\end{equation}
from which it follows by an elementary calculation that $\norm{R}\leq
\sqrt{2}$.  Thus, for any state $D$, $\beta(D)\leq \sqrt{2}$ since
$\abs{\trace{DR}}\leq \norm{R}$.  Moreover, $\beta (D)\geq 1$, since we may
always take
$A_{i}=B_{i}=I$.

If any of the four operators $A_{i},B_{i}$ is $\pm I$, then
(\ref{eq:landau}) entails that
$\norm{R}^{2}=\norm{R^{2}}=1$ and $R$ cannot indicate any CHSH
violation.  Thus, we will find it convenient to define $\gamma (D)$ in
analogy to
the definition of $\beta (D)$, but with the added restriction that the
supremum be taken over all Bell operators constructed from
\emph{nontrivial}
(i.e., not $\pm I$) self-adjoint unitary operators.  It immediately
follows that for any $D\in \alg{T}$, $\gamma (D) \in [0,\sqrt{2}]$ and
\begin{equation}
\beta (D) =\max \{ 1,\gamma (D) \} .\label{beta-gamma} \end{equation}  Thus,
any nonclassical CHSH violation indicated by $\beta (D)>1$ is
indicated just as well by $\gamma (D)>1$.

Let $D,D'\in \alg{T}$ be such that $\trnorm{D-D'}\leq \epsilon$.  Then,
for any Bell operator $R\in \alg{B}(\hil{H}_{1}\otimes \hil{H}_{2})$, it follows
from~(\ref{norms}) that \begin{equation} \abs{ \trace{DR} -
    \trace{D'R} }\leq \epsilon \norm{R} .\end{equation} In particular,
since for any Bell operator $R$, $\norm{R}\leq \sqrt{2}$,
\begin{equation} \label{eq:dr} \abs{ \trace{DR} } \leq \epsilon \sqrt{2} +\abs{
    \trace{D'R} } .\end{equation} Taking the supremum over
nontrivial Bell operators $R$, first on the right-hand side of
(\ref{eq:dr}), and then on the left, we see that
$\gamma (D)\leq \epsilon \sqrt{2}+\gamma (D')
$.  By symmetry, we have $\gamma (D')\leq \epsilon \sqrt{2}+\gamma
(D)$, so that \begin{equation} \abs{ \gamma (D)-\gamma (D') }\leq
  \epsilon \sqrt{2} \label{gamma-cont} \end{equation} and $\gamma $
is a continuous function from $\alg{T}$ (in trace norm) into
$[0,\sqrt{2}]$.  It then follows from~(\ref{beta-gamma}) that $\beta$
is a continuous function from $\alg{T}$ into $[1,\sqrt{2}]$.  Since
the set of CHSH violating density operators is the pre-image of
$(1,\sqrt{2}]$ under $\beta$, this set is open in the trace norm
topology.

Suppose now that $D$ is a convex combination $D=(1-\lambda )W+\lambda W'$ where
$W,W'\in \alg{T}$.  Then, for any Bell operator $R$,
\begin{eqnarray}
\abs{\trace{DR}}&=&\abs{ (1-\lambda)\trace{WR}+\lambda \trace{W'R} }
\nonumber \\
&\leq & (1-\lambda )\abs{\trace{WR}}+\lambda \abs{\trace{W'R}} \label{convex}
.\end{eqnarray}
Taking the supremum over nontrivial Bell operators first on the
right-hand side of~(\ref{convex}), and
then on left, we may conclude that \begin{equation}
\gamma (D)\leq (1-\lambda)\gamma (W)+\lambda \gamma (W') .\end{equation}
Thus, $\gamma$ is a convex function.  It is easy to check that
$\gamma (D)\leq 1$ for all product states
$D$, and therefore the same holds for any separable state,
by continuity and convexity of $\gamma$.

It follows from the work of Werner~\cite{wern} that when
$d_{1}=d_{2}=n\geq 2$, there are \emph{non}separable states that satisfy all
CHSH inequalities. In the case where $d_{1}=d_{2}=2$,
the Werner state, which we shall denote by $W_{22}$,
can be written as \begin{equation} \label{eq:twelve}
  W_{22}=\frac{1}{8}(I\otimes I)+\frac{1}{4}\Bigl[ (I\otimes I)-U \Bigr]
  ,\end{equation} where $U$ is the (self-adjoint, unitary) permutation
operator.
  Werner observed that for any separable density operator
$D$, we must have $\trace{UD}\geq 0$.  However, using the fact that
$U^{2}=I$ and $\trace{U}=2$, we have \begin{equation} \trace{UW_{22}}=
  \frac{1}{8}\trace{U}+\frac{1}{4}\trace{U-I}=-\frac{1}{4}<0
  .\end{equation} Thus, $W_{22}$ is nonseparable.  Moreover, using the fact
that $U=I\otimes I-2P_{s}$, where $P_{s}$ is the projection onto the
singlet state, we
may conveniently rewrite $W_{22}$ in the form: \begin{equation}
  W_{22}=\frac{1}{8}(I\otimes I)+\frac{1}{2}P_{s} .\end{equation} Since $\gamma
[(1/4)(I\otimes I)]=0$, and $\gamma$ is convex, \begin{equation}
  \gamma (W_{22})\leq \frac{1}{2}\gamma (P_{s}) =2^{-1/2}<1 ,\end{equation} and
$W_{22}$ is not CHSH violating.

More generally, we define a state $D$ to be \emph{CHSH insensitive}
whenever $D$ is nonseparable yet not CHSH violating, i.e., $\gamma(D)\leq 1$.
Such states may still violate Bell inequalities involving
projective measurements of observables with spectral values lying
outside $[-1,1]$, or more than two pairs of projective measurements, or positive
operator valued measurements.   They may also contain
``hidden'' CHSH violations in the sense that they may
violate \emph{generalized} CHSH inequalities which involve
performing consecutive projective measurements on each of the two subsystems.
To make this precise, let $\hil{H}$
be an arbitrary Hilbert space, and let $\alg{T}(\hil{H})$ be the set of density
operators on $\hil{H}$.  For any $D\in \alg{T}(\hil{H})$ and
$A\in \alg{B}(\hil{H})$ such that $ADA^{*}\not =0$, we may
define the new density operator $D^{A}$ by
\begin{equation} \label{primas}
D^{A}\equiv \frac{ADA^{*}}{\trace{ADA^{*}}}.
\end{equation}
Then $D\in \alg{T}(\hil{H}_{1}\otimes \hil{H}_{2}) (\equiv \alg{T})$
will violate a generalized CHSH inequality just in case there are
projections $Q_{1}$ and $Q_{2}$ such that $D^{Q_{1}\otimes Q_{2}}$ is
CHSH violating.  (In such a case, the violation is `seen' after
first performing a pair of selective measurements on the component
systems.)  For example, Popescu \cite{pop,merm} has shown that when
$n\geq 5$, the states constructed by Werner violate generalized CHSH
inequalities.  On the other hand, it is clear from (\ref{eq:twelve})
that $W_{22}$ itself cannot violate a generalized CHSH inequality,
since for nontrivial $Q_{1}$ or $Q_{2}$,
$W_{22}^{Q_{1}\otimes Q_{2}}$ is always a product state.

A state which violates \emph{any} Bell inequality, including
generalized inequalities, must be nonseparable.  Moreover, since the
correlations in such states---whether or not they are CHSH
sensitive---cannot be
reproduced by any local hidden
variable theory, one is justified in terming them \emph{non-local}
states.

For example, while Werner has shown that the correlations
dictated by $W_{22}$ between the outcomes of projective measurements
admit a local hidden variable model, this does not imply that $W_{22}$
is non-local; for he left it as a
conjecture
that the same is true for positive operator valued measurements
\cite[p. 4280]{wern}.

\section{CHSH Violation and Infinite-Dimensional Systems}
In this section, we establish that a bipartite system has a dense set
of non-local states when either component is infinite-dimensional.

We begin with an elementary observation about the action of
$A$ on $D$ defined by~(\ref{primas}).  This action
 is a natural generalization of the action of an
operator on unit vectors.  Indeed, we may always add an ancillary
Hilbert space $\hil{K}$ onto
$\hil{H}$ (with $\dim \hil{K}\geq \dim \hil{H}$) such that $D$ is the
reduced density operator for a pure vector
state  $x\in \hil{H}\otimes \hil{K}$.
In such a case, a
straightforward verification shows that
(when $(A\otimes I)x\neq 0$) the reduced density
operator for $(A\otimes I)x/\norm{(A\otimes I)x}$ is just $D^{A}$.

Let $\Phi$ be the map that assigns a unit vector
$x\in \hil{H}\otimes \hil{K}$ its reduced
density operator $\Phi(x)$ on $\hil{H}$.  It is easy to
see that $\Phi$ is trace-norm
continuous \cite{clif}.  Let $\{ P_{n} \}$ be any increasing sequence of
projections in $\alg{B}(\hil{H})$ with least upper bound $I$.  Then,
$(P_{n}\otimes
I)x\rightarrow x$ and \begin{eqnarray}
  D^{P_{n}}&=&\Phi [(P_{n}\otimes I)x/\norm{(P_{n}\otimes I)x}] \\
  &\rightarrow &\Phi [x] =D , \end{eqnarray} where the convergence is
in trace norm. We make use of this convergence in our
arguments below.

{\bf Proposition 1.}  \emph{If $d_{1}=d_{2}=\infty$, then the set of CHSH
violating states is trace norm dense in the set of all
density operators on $\hil{H}_{1}\otimes \hil{H}_{2}$.}

\emph{Proof:}
Fix an arbitrary density operator $D$ on $\hil{H}_{1}\otimes
  \hil{H}_{2}$, and fix orthonormal bases for the factor
  spaces $\{e_{i}\}$ and $\{f_{j}\}$.
  Let $P_{n}$ be the projection onto the span of
  $\{e_{i}\otimes f_{j}\}_{i,j\leq n}$, and set
  \begin{equation}
\psi_{n}=\frac{1}{\sqrt{2}}(\ket{e_{n+1}}\ket{f_{n+1}}+\ket{e_{n+2}}\ket{f_{
n+2}}).
  \end{equation}
  Consider the sequence of density operators
  $\{D_{n}\}$ defined by
  \begin{equation} \label{eq:dops}
  D_{n}=(1-\frac{1}{n})D^{P_{n}}+\frac{1}{n}P_{\psi_{n}}
  \end{equation}
  where $P_{\psi}$ projects onto the ray $\psi$ generates.  Since
  $\lim_{n\rightarrow \infty}D_{n}=D$ in trace norm, all that remains
  to show is that each $D_{n}$ is CHSH violating.  As $\psi_{n}$ is
  the pure singlet state, there are ``spin components'' (i.e.
  self-adjoint unitaries) $A^{n}_{i}$, $B^{n}_{i}$ ($i=1,2$) such that
  each $A^{n}_{i}$ leaves the subspace generated by
  $\ket{e_{n+1}},\ket{e_{n+2}}$ invariant and acts like the identity
  on the complement; similarly for each $B^{n}_{i}$ and the subspace
  generated by $\ket{f_{n+1}},\ket{f_{n+2}}$; and, moreover, the Bell
  operator
  \begin{equation}
  R_{n}\equiv\frac{1}{2}(A^{n}_{1}\otimes B^{n}_{1}+ A^{n}_{1}\otimes
  B^{n}_{2}+A^{n}_{2}\otimes B^{n}_{1}-A^{n}_{2}\otimes B^{n}_{2})
  \end{equation}
  is such that $\trace{P_{\psi_{n}}R_{n}}>1$.  Therefore, in view
  of~(\ref{eq:dops}), to show that $\trace{D_{n}R_{n}}>1$, and hence
  that $D_{n}$ is CHSH violating, it suffices to observe that
  $\trace{D^{P_{n}}R_{n}}= 1$.  But this is immediate from the fact
  that $R_{n}$ acts as the
  identity on $P_{n}$'s range.  \mbox{\emph{QED}}

A similar argument shows that non-local states are dense in the
case $d_{1} < d_{2} = \infty$.

{\bf Proposition 2.}  \emph{If $d_{1}<d_{2}=\infty$, then the set of
non-local states is trace norm dense in the set of all
density operators on $\hil{H}_{1}\otimes \hil{H}_{2}$.}

\emph{Proof:} Fix an arbitrary density operator $D$ on
$\hil{H}_{1}\otimes \hil{H}_{2}$, and fix orthonormal bases for the
factor spaces $\{e_{i}\}_{i=1}^{d_1}$ and $\{f_{j}\}_{j=1}^{\infty}$.
Let $P'_{n}$ be the projection onto the span of $\{e_{i}\otimes
f_{j}\}_{1 \leq i \leq d_1, 1 \leq j \leq n}$, and set \begin{equation}
\psi'_{n}=\frac{1}{\sqrt{2}}(\ket{e_{1}}\ket{f_{n+1}}+\ket{e_{2}}\ket{f_{n+2}}).
  \end{equation}
  Consider the sequence of density operators
  $\{D_{n}\}$ defined by
  \begin{equation} \label{eq:dn}
  D_{n}=(1-\frac{1}{n})D^{P'_{n}}+\frac{1}{n}P_{\psi'_{n}} .
  \end{equation}
  As before, $\lim_{n\rightarrow \infty}D_{n}=D$ in trace norm, so it
  suffices to show that each $D_{n}$ is non-local.
  Define the projections $Q_1 , Q_2$ onto the spans of $\{ e_{i} \}_{1
    \leq i \leq 2}$ and $\{ f_{j} \}_{n+1 \leq j \leq n+2 }$,
  respectively.  Then since $D_{n}^{Q_1 \otimes Q_2}=P_{\psi'_{n}}$,
  $D_{n}$ violates a generalized CHSH inequality.  \ \mbox{\emph{QED}}

  Note that Prop. 2 entails that when
  $d_{2}=\infty$, the set of nonseparable states is dense.  This
  reproduces, by quite different methods,  the main result of
  \cite{clif}.

\section{Generic CHSH Violation Characterizes Infinite-Dimensional Systems}
As mentioned in the introduction, when both $d_{1},d_{2}<\infty$,
there is always an open neighborhood of separable states
\cite{horo,caves,zguy}.  Since separable states cannot display any
nonlocal correlations, it follows that in this case the CHSH violating
states cannot be dense.  Note, however, that this same method of
argument could not establish an open CHSH non-violating neighborhood
in the case where $d_{1}<d_{2}=\infty$, for in that case we know that
the separable states are nowhere dense.  However, as we now show,
such neighborhoods exist.

Let $D\in \alg{T}$ be a density operator with $\gamma (D)<1$.
It is not difficult to see that the distance from $D$ to the set of
CHSH violating states is bounded below by $2^{-1/2}(1-\gamma
(D))$.  Indeed, for any density operator $D'$, if \begin{equation}
\trnorm{D-D'}\leq 2^{-1/2}(1-\gamma (D)) ,\end{equation} then
from~(\ref{gamma-cont}), \begin{equation} \gamma (D')\leq 2^{1/2}
\Bigl[ 2^{-1/2}(1-\gamma (D)) \Bigr] +\gamma (D) =1 .\end{equation}
Thus any state $D$ with $\gamma (D)<1$ is
surrounded by a neighborhood of states that are again not CHSH
violators.

{\bf Proposition 3.} \emph{If $d_{1}<\infty$ then, for any density operator
$D_{2}\in
\alg{T}(\hil{H}_{2})$, we have} \begin{equation}
\gamma [d_{1}^{-1}(I\otimes D_{2})] \leq 1-2d_{1}^{-1}<1 .\end{equation}

\emph{Proof:} Let $A$ be a self-adjoint unitary operator (not $\pm I$)
acting on $\hil{H}_{1}$.  Then $A=P_{1}-P_{2}$, where $P_{i}$ is a
projection $(i=1,2)$.  Since $A\neq \pm I$, $P_{1}\neq 0$ and
$P_{2}\neq 0$.  Thus,
\begin{eqnarray} \abs{ \trace{d_{1}^{-1}A}
    }&=& d_{1}^{-1}\biggl| \trace{P_{1}}-\trace{P_{2}} \biggr| \\
  &\leq & d_{1}^{-1}(d_{1}-2) =1-2d_{1}^{-1} \label{smaller}. \end{eqnarray}
Now let $R$ be any Bell-operator for $\hil{H}_{1}\otimes \hil{H}_{2}$,
constructed from (nontrivial) self-adjoint unitary operators.  Then,
\begin{eqnarray}
\lefteqn{\abs{\trace{d_{1}^{-1}(I\otimes D_{2})R}} }\qquad \qquad && \\
&=& \frac{1}{2}\biggl| \trace{d_{1}^{-1}(A_{1}+A_{2})}\cdot
\trace{D_{2}B_{1}} \nonumber \\
& &+\trace{d_{1}^{-1}(A_{1}-A_{2})}\cdot \trace{D_{2}B_{2}} \biggr|  \\
&\leq &\frac{1}{2}\biggl| \trace{d_{1}^{-1}A_{1} }+\trace{d_{1}^{-1}A_{2} }
\biggr| \nonumber \\
& &+\frac{1}{2}\biggl| \trace{d_{1}^{-1}A_{1} } -\trace{d_{1}^{-1}A_{2} }
\biggr|  \\
&\leq & 1-2d_{1}^{-1} .\end{eqnarray}
The last inequality follows since \begin{equation}
  \abs{a_{1}+a_{2}}+\abs{a_{1}-a_{2}} \leq 2 \max \{ \abs{a_{i}} \}
  \label{easy} ,\end{equation} for any two real numbers $a_{1},a_{2}$.
\emph{QED}

Note that the considerations prior to this proposition entail that
$d_{1}^{-1}(I\otimes D_{2})$ lies in a neighborhood of CHSH
non-violating states of (trace norm) size at least
$d_{1}^{-1}\sqrt{2}$.  (Of course, this estimate could be improved if
restrictions on $D_{2}$ were also taken into account.)

{\bf Proposition 4.}  \emph{The set of CHSH violating density operators is
trace norm dense in the set of all density operators on $\hil{H}_{1}\otimes
\hil{H}_{2}$ (and its complement is nowhere dense) if and only if $\dimn
\hil{H}_{1}=\dimn \hil{H}_{2}=\infty$.}

\emph{Proof:} Suppose that $d_{1}=d_{2}=\infty$.  Then, from Prop.~1,
the set of CHSH violating states is trace norm dense (and its
\emph{closed} complement must be nowhere dense).  Conversely, if
$d_{1}<\infty$, then Prop.~3 (and the discussion preceding it) ensures
the existence of many open neighborhoods of states that satisfy
the CHSH inequality.  \emph{QED}

\section{CHSH insensitive states}
Props. 1--4 establish that CHSH insensitive states exist in the
case when $d_{1}<d_{2}=\infty$.  In particular, since there is at
least one open set of states that do not violate the CHSH inequality,
and since the nonseparable states are dense, there must be
nonseparable states that are not CHSH violating.  Indeed, Prop.~3
provides us with a class of states which we know have a surrounding
neighborhood of states that are not CHSH violating, while Prop.~2
shows how, given any state, we may construct a sequence of
nonseparable states which converges to that state.  In Appendix B, we
invoke the alternate method of constructing nonseparable states given in
\cite{clif} to construct
a sequence of CHSH insensitive states that
converges continuously to a product state.  (We do so only for the simplest
case of a bipartite
system with exactly one two-dimensional component---such as a
spin-1/2 particle, distinguishing its internal and
external degrees of freedom.)

 We have not so far shown that there are CHSH insensitive states in the
cases $d_{1}< d_2 <
\infty$ and $d_{1}=d_{2}=\infty$.  We now proceed to show that in all
relevant cases, i.e.,
whenever $d_{1},d_{2}\geq 2$, CHSH insensitive states exist.  Moreover, if
$d_{1}<\infty$,
there is always an open neighborhood of CHSH insensitive states.

CHSH insensitive states can be constructed simply by embedding the $2\times
2$ Werner state
$W_{22}$ into the higher-dimensional space.  Let $\{ e_{i}\otimes f_{j} \}$
denote an
orthonormal product basis for $\hil{H}_{1}\otimes \hil{H}_{2}$, and let
$\hil{K}$ denote the
$2\times 2$ subspace spanned by $\{ e_{i}\otimes f_{j}:i,j=1,2 \}$.  Note
that the projection
onto $\hil{K}$ is just the product $P\otimes Q$ of the projections $P$ onto
$\{ e_{i}:i=1,2 \}$ and $Q$ onto $\{ f_{j}:j=1,2 \}$.  Corresponding
to the permutation operator $U$ of $\mbox{$\mathbb{C}$}^{2}\otimes
\mbox{$\mathbb{C}$}^{2}$, we let $U'$ denote the (partial isometry)
operator on $\hil{H}_{1}\otimes \hil{H}_{2}$ which permutes the basis
elements of $\hil{K}$ and maps $\hil{K}^{\perp}$ to $0$.  Then, by
analogy with $W_{22}$, we may
define \begin{equation} W_{22}'\equiv \frac{1}{8}(P\otimes
  Q)+\frac{1}{4}\Bigl[ (P\otimes Q)-U' \Bigr] .
\end{equation} It is not difficult to see that
$W_{22}'\in \alg{T}(\hil{H}_{1}\otimes \hil{H}_{2})$. We now verify that
$W_{22}'$, \emph{as a state of} $\hil{H}_{1}\otimes
\hil{H}_{2}$, is again CHSH insensitive.

For a density operator $D\in\alg{T}(\hil{H}_{1}\otimes \hil{H}_{2})$,
let us say that $D$ is $\hil{K}$-\emph{separable} just in case $D$ is
in the closed convex hull of product states \emph{all of whose ranges
  are contained in $\hil{K}$}.

{\bf Proposition 5.}  \emph{Suppose that $D\in \alg{T}(\hil{H}_{1}\otimes
\hil{H}_{2})$ and $D$ has range contained in $\hil{K}$.  If $D$ is
separable, then $D$ is $\hil{K}$-separable.}

Before we give the proof of this proposition, we recall from
\cite{clif} some basic
facts concerning the operation $D\rightarrow D^{A}$ on density
operators defined in~(\ref{primas}).  Suppose that $D\in
\alg{T}(\hil{H})$ is a convex combination of density
operators \begin{equation}
D=\sum _{i=1}^{n}\lambda _{i} D_{i} . \end{equation}
Then, for any $A\in \alg{B}(\hil{H})$, if $ADA^{*}\neq 0$, we may set
\begin{equation} \lambda ^{A}_{i}\equiv \lambda _{i}
\frac{\trace{AD_{i}A^{*}}}{\trace{ADA^{*}}} , \label{eq:hi} \end{equation}
and we have
\begin{equation} D^{A}\equiv \frac{ADA^{*}}{\trace{ADA^{*}}}=\sum
_{i=1}^{n}\lambda ^{A}_{i} D ^{A}_{i} , \label{eq:ho} \end{equation}
where $\sum _{i=1}^{n}\lambda ^{A}_{i}=1$.  Thus, $D^{A}$ is a convex
combination of the $D^{A}_{i}$.  Note, also, that when $ADA^{*}\neq 0$, the
operation
$D\rightarrow D^{A}$ is trace norm continuous at $D$ (since multiplication
by a fixed
element in $\alg{B}(\hil{H})$ is trace norm continuous~\cite[p.~39]{schat}.)

Further specializing to the case where $\hil{H}\equiv \hil{H}_{1}\otimes
\hil{H}_{2}$, note that if $D=D_{1}\otimes D_{2}$ is a product state,
and $A\in \alg{B}(\hil{H}_{1})$, $B\in \alg{B}(\hil{H}_{2})$ are
arbitrary, then \begin{equation} D^{A\otimes B}=(D_{1}\otimes
  D_{2})^{A\otimes B}=D_{1}^{A}\otimes D_{2}^{B} .\end{equation}

\emph{Proof of the proposition:} Let $P\otimes Q$ denote the
projection onto $\hil{K}$.  Since $D$ has range contained in
$\hil{K}$, we have $D^{P\otimes Q}=D$.  Suppose now that $D$ is
separable.  That is, $D=\lim _{n}D_{n}$ where each $D_{n}\in
\alg{T}(\hil{H}_{1}\otimes \hil{H}_{2})$ is a convex combination of
product states.  Thus, by continuity we have \begin{equation}
  D=D^{P\otimes Q}=\lim _{n}D_{n}^{P\otimes Q} .\end{equation}
By the preceding considerations, each $D_{n}^{P\otimes Q}$ is
a convex combination of product states, each of which has range
contained in $\hil{K}$.  Thus, $D$ is $\hil{K}$-separable.  \emph{QED}

It is now straightforward to see that $W_{22}'$ is nonseparable.
Indeed, since $W_{22}'$ has range contained in $\hil{K}$, if $W_{22}'$
were separable, it would also be $\hil{K}$-separable.  However, using
the natural isomorphism between $\mbox{$\mathbb{C}$}^{2}\otimes
\mbox{$\mathbb{C}$}^{2}$ and
$\hil{K}$, and the induced isomorphism between density operators on
$\mbox{$\mathbb{C}$}^{2}\otimes \mbox{$\mathbb{C}$}^{2}$ and density
operators on $\hil{H}_{1}\otimes \hil{H}_{2}$ with range in $\hil{K}$,
it would follow that $W_{22}$ is separable.  Therefore, $W_{22}'$ is
nonseparable.

To see that $W_{22}'$ is not CHSH violating, note that for any Bell
operator $R$ for $\hil{H}_{1}\otimes \hil{H}_{2}$, \begin{equation}
  R'\equiv (P\otimes Q)R(P\otimes Q) \end{equation} is again a
Bell operator (constructed out of self-adjoint contractions
$PA_{i}P,QB_{i}Q$ that may not be
unitary).  Moreover, \begin{eqnarray}
  \abs{ \trace{W_{22}'R} }&=&\abs{ \trace{(W_{22}')^{P\otimes Q}R}} \\
  &=& \abs{\trace{W_{22}'(P\otimes Q)R(P\otimes Q)}} \\
  &=& \abs{\trace{W_{22}'R'}}.  \end{eqnarray} Thus, if $W_{22}'$
  violates a CHSH inequality, it must violate a CHSH inequality
 with respect to some Bell operator $R'$
whose range lies in $\hil{K}$.  But any such $R'$ has a counterpart
in $\alg{B}(\mbox{$\mathbb{C}$}^{2}\otimes \mbox{$\mathbb{C}$}^{2})$
that would display a CHSH violation for $W_{22}$.  Therefore,
$W_{22}'$ is not CHSH violating and is CHSH insensitive for $\hil{H}_{1}\otimes
\hil{H}_{2}$.

We end by combining the fact that there are always CHSH insensitive states
with the results of the previous section to show that there are ``many''
CHSH insensitive states,
unless both component spaces are infinite-dimensional.

{\bf Proposition 6.}  \emph{There is an open set of CHSH insensitive density
 operators on $\hil{H}_{1}\otimes \hil{H}_{2}$ if and only if $\dimn
 \hil{H}_{1}<\infty$ or $\dimn \hil{H}_{2}<\infty$.}

\emph{Proof:} The ``only if'' follows immediately from Prop. 1.
(If $d_{1}=d_{2}=\infty$, then the CHSH insensitive states are
contained in the nowhere dense set of states which satisfy all CHSH
inequalities.)  To prove the converse, suppose that $d_{1}<\infty$.
It would suffice to show that there is a nonseparable state $W$ with
$\gamma (W)<1$.  For, in that case, we may use the continuity of
$\gamma$ to obtain an open neighborhood $\mathcal{O}$ of $W$ which
contains only states with no CHSH violations.  Taking the intersection
of $\mathcal{O}$ with the open set of nonseparable states would give
the desired open set of CHSH insensitive states.

  From considerations adduced above, there is always a CHSH
insensitive state $W'\in \alg{T}(\hil{H}_{1}\otimes \hil{H}_{2})$.
Since $W'$ does not violate a CHSH inequality, we have $\gamma
(W')\leq 1$.  Moreover, from Prop.~3, there is a $D\in
\alg{T}$ such that $\gamma (D)<1$.  For each $n$, let \begin{equation}
  W_{n}\equiv (1-n^{-1})W'+n^{-1}D .\end{equation} Clearly,
$W_{n}\rightarrow W'$ in trace norm, and by the convexity of $\gamma$,
\begin{eqnarray} \gamma (W_{n})&\leq & ( 1-n^{-1}) \gamma (W')+n^{-1}\gamma
(D) \\
&\leq & ( 1-n^{-1} ) +n^{-1} \gamma (D) <1  ,\end{eqnarray}
for all $n$.  However, since $W'$ is nonseparable, and the nonseparable
states are open, there is an $m\in \mathbb{N}$ such that $W_{n}$ is
nonseparable for all $n\geq m$.  Thus, setting $W\equiv W_{m}$ gives
the desired nonseparable state with $\gamma (W)<1$.  \emph{QED}

\section{Conclusion}
We have established the conjecture made in~\cite{clif} that bipartite
systems whose components are both infinite-dimensional (e.g., a pair of
particles, neglecting their spins) have states that generically
violate the CHSH inequality.  We also established that even if one of
the components is finite-dimensional  (e.g., a
spin-1/2 particle), non-locally correlated states
remain dense.   Finally, we have identified new classes of
CHSH insensitive states for finite by infinite systems, and established
that such states can only be
neglected, for all practical purposes, in the infinite by infinite
case.

Infinite-dimensional systems thus provide a
resource of nonlocality which --- practically speaking ---
cannot be completely destroyed
by noise or by errors in preparation or measurement.
In this they differ from finite-dimensional systems, where
entangled mixed states can always be reduced to separable
states by sufficient noise.
One might naively conclude that, to the extent that it is practicable
in quantum information and computation theory to exploit
infinite-dimensional systems, it would be advantageous to
do so.  But in fact we can never
exploit all the degrees of freedom in a infinite-dimensional
system.  So, though we hope the above results may be useful in
developing the theory of entanglement in large finite-dimensional systems,
we doubt that they themselves can lead to direct practical application.

Even in the case of large finite-dimensional systems, there
is a potential pitfall.  It may well be that nonlocality becomes
harder and harder to destroy, by some sensible quantitative
measure, as the size of the system becomes larger.  However,
the nonlocality results we have outlined give no
indication of a general procedure for extracting or demonstrating
nonlocality.  Protecting some form of nonlocality is less useful
if it is achieved at the cost of making it harder and harder
to find.  It would thus be very interesting to quantify the
trade-offs which can usefully be made in this direction when large
finite-dimensional systems are used to counter noise on a highly
noisy channel.

\appendix
\section{} \label{app}
We give here a self-contained version of Summers and Werner's
argument\cite{summ}
that $\beta (D)$ is equal to the supremum of
$\abs{ \trace{DR} }$, where $R$ only runs over the Bell operators for
$\hil{H}_{1}\otimes \hil{H}_{2}$ that are constructed from
self-adjoint \emph{unitary} operators.

Recall that the weak-operator topology on $\alg{B}(\hil{H})$ is the
coarsest topology for which all functionals of the form
\begin{equation} T\rightarrow \abs{ \langle Tx,y \rangle } \qquad
  \qquad x,y\in \hil{H} ,\end{equation} are continuous at $0$.  It
then follows that the unit ball of $\alg{B}(\hil{H})$ is compact in
the weak-operator topology~\cite[Thm.~5.1.3]{kad}.  (Of course, if
$\dimn\ \hil{H}< \infty$, the unit ball of $\alg{B}(\hil{H})$ is also
compact in the operator-norm topology.)  Moreover, since the adjoint
operation is weak-operator continuous, the set of self-adjoint
operators is weak-operator closed in $\alg{B}(\hil{H})$, and
$\alg{B}(\hil{H})_{s}$ is weak-operator compact (as well as convex).

Fix $A_{2}\in \alg{B}(\hil{H}_{1})_{s}$ and $B_{1},B_{2}\in
\alg{B}(\hil{H}_{2})_{s}$.
We show that the map $\Psi _{D}:\alg{B}(\hil{H}_{1})_{s}\rightarrow
\mathbb{R}$ defined by
\begin{equation}
\Psi_{D} (A_{1}) \equiv \frac{1}{2}\mbox{$\mathrm{Tr}$} \Bigl( D( A_{1}\otimes
(B_{1}+B_{2}) \mbox{}+ A_{2}\otimes (B_{1}-B_{2}) ) \Bigr)
,\end{equation}
is affine and weak-operator continuous.  From this it will follow that
$\Psi_{D}$ attains
its extremal values on extreme points of
$\alg{B}(\hil{H}_{1})_{s}$~\cite[Prop.~7.9]{conw}.
These, however, consist precisely of the self-adjoint
unitary operators~\cite[Prop.~7.4.6]{kad}.

Now, to establish that $\Psi_{D}$ is affine and weak-operator
continuous, let $\Lambda_{D} :\alg{B}(\hil{H}_{1})_{s}\rightarrow
\mathbb{R}$ denote the linear functional defined by
\begin{equation}
\Lambda _{D} (A_{1})\equiv \mbox{$\mathrm{Tr}$} \Bigl[ D(A_{1}\otimes
(1/2)(B_{1}+B_{2}))
\Bigr] . \end{equation}
Then, $\Lambda_{D}$ is the composition of the map \begin{equation}
A_{1}\rightarrow A_{1}\otimes (1/2)(B_{1}+B_{2}) ,\end{equation} from
$\alg{B}(\hil{H}_{1})_{s}$ into $\alg{B}(\hil{H}_{1}\otimes
\hil{H}_{2})_{s}$, with the
functional $\trace{D~\cdot ~}$.  However, the former is continuous (when
both algebras are
equipped with the weak-operator topology) since multiplication by a fixed
operator is weakly
continuous.  Moreover, $\trace{D~\cdot ~}$ is weak-operator continuous on
the unit ball of
$\alg{B}(\hil{H}_{1}\otimes \hil{H}_{2})$.  Thus, $\Lambda_{D}$ is
weak-operator continuous.
Now, let
\[ r_{D}\equiv \mbox{$\mathrm{Tr}$} \Bigl[ D(A_{2}\otimes
(1/2)(B_{1}-B_{2}) ) \Bigr] . \]
Then, $\Psi_{D}=\Lambda _{D} +r_{D}$ is affine and weak-operator continuous.

  From the above considerations it now follows that for every $A_{1}\in
\alg{B}(\hil{H}_{1})_{s}$ and Bell operator $R$ constructed using
$A_{1}$, there is a Bell operator $R'$ constructed from the same
elements as $R$, except with $A_{1}$ replaced by a self-adjoint
unitary operator, and such that $\abs{\trace{DR}}\leq
\abs{\trace{DR'}}$.  By symmetry, the same conclusion applies to
$A_{2},B_{1}$ and $B_{2}$.  Thus, for any given Bell operator $R$,
there is a Bell operator $R'$ constructed entirely from self-adjoint
unitaries, and such that $\abs{\trace{DR}}\leq \abs{\trace{DR'}}$.

\section{}
In this appendix, we use the results of the current paper and
of~\cite{clif} to construct a
continuous ``path'' of CHSH insensitive states with endpoint
a product state.  Reversing the
convention $d_{1}\leq d_{2}$ of the current paper (to align with that chosen in
\cite{clif}), we examine the case where
$d_{1}=\infty$ and $d_{2}=2$.

Let
$\{ e_{i}\}\subseteq\hil{H}_{1}$ and $\{ f_{1},f_{2} \}\subseteq\hil{H}_{2}$
be orthonormal bases.  Attaching an ancillary Hilbert space $\hil{H}_{3}$,
with infinite
orthonormal basis $\{ g_{k}\}$, we may define a unit
vector $v_{0}\in \hil{H}_{1}\otimes \hil{H}_{2}\otimes \hil{H}_{3}$ by
\begin{eqnarray} v_{0} &\equiv & \frac{1}{2}\Bigl
  ( \ket{e_{1}}\ket{f_{1}}\ket{g_{1}}
+ \ket{e_{2}}\ket{f_{2}}\ket{g_{2}} \nonumber \\
& &\mbox{}  +\ket{e_{2}}\ket{f_{1}}\ket{g_{3}}+
\ket{e_{1}}\ket{f_{2}}\ket{g_{4}} \Bigr) .\end{eqnarray}
Note that the reduced density operator $\Phi
(v_{0})\in \alg{T}(\hil{H}_{1}\otimes \hil{H}_{2})$ for $v_{0}$ is just
$\frac{1}{2}P\otimes \frac{1}{2}I$, where $P$ is the projection onto
the subspace of $\hil{H}_{1}$ spanned by $\{ e_{1},e_{2} \}$.  Thus,
from Prop.~3 (interchanging $1$ with $2$), there is a CHSH
non-violating neighborhood surrounding $\Phi (v)$.

Now, for each $\lambda \in [0,1]$, define the unit vector $v_{\lambda
  }\in\hil{H}_{1}\otimes \hil{H}_{2}\otimes \hil{H}_{3}$ by
\begin{equation}
  v_{\lambda }\equiv(1-\lambda) v_{0} +[\lambda(2-\lambda)]^{1/2}u
  \end{equation}
  where $u$ is the unit vector
  \begin{equation}
  u\equiv  \sum _{n=1}^{\infty}
2^{-(n+1)/2}\Bigl(\ket{e_{2n+1}}\ket{f_{1}}\ket{g_{n}}
  +\ket{e_{2n+2}}\ket{f_{2}}\ket{g_{n}}\Bigr) .\end{equation}
Clearly, $v_{\lambda }\rightarrow v_{0}$ as $\lambda \rightarrow 0$.
Furthermore, by the continuity of $\Phi$, $\Phi (v_{\lambda
  })\rightarrow
\Phi (v_{0}) $. It then follows that there is an $\epsilon >0$ such
  that $\Phi (v_{\lambda })$ is not CHSH violating for all $\lambda
<\epsilon $.  However, by
construction $v_{\lambda }$ is \emph{separating} for the subalgebra $I\otimes
\alg{B}(\hil{H}_{2}\otimes \hil{H}_{3})$, for all $\lambda \in
  (0,1]$.  That is, for any $A\in \alg{B}(\hil{H}_{2}\otimes
  \hil{H}_{3})$, if $(I\otimes A)v_{\lambda}=0$, then
  $A=0$. (To see this, observe that any such $A$ would have to annihilate
all the basis
  vectors $\{f_{j}\otimes g_{k}\}$ due to the orthogonality of
  the $\{e_{i}\}$.)  Thus, invoking \cite[Lemmas~1,2]{clif},
each $\Phi (v_{\lambda })$ is nonseparable, and, for all
$0<\lambda<\epsilon$, CHSH insensitive.

\
\vskip10pt
\centerline{\bf Acknowledgements}
\vskip5pt
A.K. thanks the Royal Society for financial support.

\end{multicols}
\end{document}